\documentclass[twocolumn]{aastex631}

\usepackage{booktabs}
\usepackage{multirow}
\usepackage{float}
\usepackage{graphicx}	% Including figure files
\usepackage{amsmath}	% Advanced maths commands
\usepackage{amssymb}	% Extra maths symbols
\usepackage{breqn}
\usepackage[para,online,flushleft]{threeparttable}

\begin{document}

\title{Exploring giant barium stars:\\$^{12}\rm{C}/^{13}\rm{C}$ ratio and elemental abundances of carbon, nitrogen, and oxygen\footnote{The program stars were observed under the programs ID 079.A-9200(A), 080.A-9206(A), 081.A-9203(A), 082.A-9200(A), 083.A-9206(A), 084.A-9205(A), and 085.A-9202(A), under agreement between Observat\'orio Nacional (Brazil) and European Southern Observatory (ESO).}}

\correspondingauthor{M. P. Roriz}
\email{michelle@on.br}
\author[0000-0001-9164-2882]{M. P. Roriz}
\affiliation{Observat\'orio Nacional/MCTI, Rua General Jos\'e Cristino, 77, 20921-400, Rio de Janeiro, Brazil}

\author[0000-0003-4842-8834]{N. A. Drake}
\affiliation{Observat\'orio Nacional/MCTI, Rua General Jos\'e Cristino, 77, 20921-400, Rio de Janeiro, Brazil}
\affiliation{Laboratory of Observational Astrophysics, Saint Petersburg State University, Universitetski pr. 28, 198504, Saint Petersburg, Russia}

\author[0000-0002-8504-6248]{N. Holanda}
\affiliation{Observat\'orio Nacional/MCTI, Rua General Jos\'e Cristino, 77, 20921-400, Rio de Janeiro, Brazil}

\author{M. Lugaro}
\affiliation{Konkoly Observatory, HUN-REN Research Centre for Astronomy and Earth Sciences, Konkoly Thege Mikl\'os út 15-17, H-1121, Hungary}
\affiliation{ELTE E\"{o}tv\"{o}s Lor\'and University, Institute of Physics, Budapest 1117, P\'azm\'any P\'eter s\'et\'any 1/A, Hungary}
\affiliation{School of Physics and Astronomy, Monash University, VIC 3800, Australia}
\affiliation{CSFK, MTA Centre of Excellence, Budapest, Konkoly Thege Mikl\'os út 15-17., H-1121, Hungary}

\author{B. Cseh}
\affiliation{Konkoly Observatory, HUN-REN Research Centre for Astronomy and Earth Sciences, Konkoly Thege Mikl\'os út 15-17, H-1121, Hungary}
\affiliation{CSFK, MTA Centre of Excellence, Budapest, Konkoly Thege Mikl\'os út 15-17., H-1121, Hungary}
\affiliation{MTA-ELTE Lend{\"u}let "Momentum" Milky Way Research Group, Hungary}

\author{S. Junqueira}
\affiliation{Observat\'orio Nacional/MCTI, Rua General Jos\'e Cristino, 77, 20921-400, Rio de Janeiro, Brazil}

\author{C. B. Pereira}
\affiliation{Observat\'orio Nacional/MCTI, Rua General Jos\'e Cristino, 77, 20921-400, Rio de Janeiro, Brazil}




\begin{abstract}
Barium (Ba) stars belong to binary systems that underwent mass transfer events. As a consequence, their envelopes were enriched with material synthesized in the interiors of their evolved companions via \textit{slow} neutron-capture nucleosynthesis, the $s$-process. As post-interacting binaries, Ba stars figure as powerful tracers of the $s$-process. In this study, we conduct a classical local thermodynamic equilibrium analysis for a sample of 180 Ba giant stars to find complementary insights for the $s$-process, in form of elemental abundances of carbon, nitrogen, and oxygen, as well as the $^{12}\rm{C}/^{13}\rm{C}$ ratio. We found carbon abundances systematically larger than those observed in normal giants, with [C/Fe] ratios ranging within from $-0.30$ to $+0.60$~dex. As expected, the [C/Fe] ratios increase for lower metallicity regimes and are strongly correlated with the average $s$-process abundances. Nitrogen abundances have a flat behavior around $\rm{[N/Fe]}\sim+0.50$~dex and are moderately correlated with sodium abundances. Except for HD~107541, the entire sample shows $\rm{C/O}<1$. We found $^{12}\rm{C}/^{13}\rm{C}<20$ for $\sim80\%$ of the sampled stars and $^{12}\rm{C}/^{13}\rm{C}>60$ for three objects.
\end{abstract}




\section{Introduction} \label{sec:intro}

Barium (Ba) stars were first identified by \cite{bidelman1951} as a class of peculiar red giants. The abnormal strengthening of Ba\,{\sc ii} and Sr\,{\sc ii} absorption features initially observed in their spectra posed a challenge to stellar evolution theory. As first ascent red giants, Ba stars are not able to internally synthesize heavy elements such as Ba and Sr. These, along with roughly half of the cosmic abundances for the elements beyond the iron peak $(Z>30)$, are mostly produced through \textit{slow} neutron captures, the $s$-process, starting on Fe-seed nuclei \citep[][]{b2fh1957, kappeler2011, lugaro2023}. The physical conditions for such a nucleosynthetic pathway are found in the deep layers of Asymptotic Giant Branch (AGB) stars, during their Thermally-Pulsing (TP-AGB) phase. The $s$-process takes place within the He-rich region located between the He- and H-burning shells \citep[][]{gallino1998, busso1999, karakas2014}. 

\cite{mcclure1980} realized the binary nature of Ba stars and hypothesized mass transfer as the origin of the anomalies observed in these objects. In this scenario, Ba stars had their atmospheres contaminated by the outflows of their evolved binary companions, which underwent the TP-AGB phase and became faint white dwarfs. The binary nature of Ba stars is widely supported by observational data \citep[][]{jorissen2019, escorza2023}, and many studies over the years based on high-resolution spectroscopy have confirmed the overabundance of other $s$-process elements in their envelopes, in addition to Ba and Sr \citep[e.g.,][]{allen2006, pereira2011b, decastro2016, karinkuzhi2018b, shejeelammal2020, roriz2021a, roriz2021b, roriz2024_gba}. Thus, as post-interacting binaries, Ba stars are ideal targets to probe the evolution of binary systems \citep[][]{escorza2020}{}{}, the physics of mass transfer \citep[][]{jorissen1998}{}{}, as well as the $s$-process nucleosynthesis \citep[][]{allen2006b, cseh2018, cseh2022, denhartogh2023, vilagos2024}{}{}.

The study of \citet[][hereafter Paper~I]{decastro2016}, in particular, concentrated efforts on a chemical and kinematics analysis for a sample of $\sim 180$ Ba giant stars. Paper~I reported abundances of Na, Al, $\alpha$-elements, iron-peak elements, and the heavy elements Y, Zr, Ce, and Nd\footnote{The La abundances reported in Paper~I were later revised and updated in Paper III.}. In the subsequent studies, we added to the initial sample the metal-rich Ba stars analyzed by \citet[][]{pereira2011b} and expanded the abundance patterns observed in these stars, reporting abundances of Rb \citep[][Paper~II]{roriz2021a}{}{}, Sr, Nb, Mo, Ru, La, Sm, Eu \citep[][Paper~III]{roriz2021b}{}{}, and more recently W \citep[][Paper~IV]{roriz2024_gba}{}{}. These observational data sets, along with machine-learning techniques, have provided strong constraints to $s$-process nucleosynthesis models of AGB stars \citep[see][]{denhartogh2023, vilagos2024}.

In addition to neutron-capture elements, chemical abundances of specific light elements, such as carbon, nitrogen, and oxygen, are of particular interest in Ba stars. Sensitive to mixing events in different stellar evolution stages, these elements can provide us with valuable insights to trace back the former TP-AGB stars that polluted the envelopes of the Ba stars. Therefore, as a continuation of a series of studies dedicated to spectral analyses of Ba stars started in Paper~I, we report in this fifth work chemical abundances of carbon, nitrogen, and oxygen, in addition to carbon isotopic ratio, $^{12}\rm{C}/^{13}\rm{C}$, for the same targets sampled in the previous papers of this series. In Section~\ref{sec:sample}, we briefly present the sampled stars; in Section~\ref{sec:abundance}, we describe the methodology employed to derive elemental abundances and the carbon isotopic ratio. In Section~\ref{sec:analysis}, we analyze and discuss the results; in Section~\ref{sec:models}, the data are examined in light of the $s$-process models. Concluding remarks are drawn in Section~\ref{sec:conclusions}.

\section{Target stars} \label{sec:sample}

As previously stated, the stars analyzed here are the same objects sampled in the previous works in this series. In summary, their spectra were obtained as a result of a spectroscopic survey conducted between 1999 and 2010, in search of new chemically peculiar stars. The instrument used to extract the stellar spectra was the Fiber-fed Extended Range Optical Spectrograph (FEROS; \citealt{kaufer1999}), attached to the 1.52 and 2.2~m ESO telescopes at La Silla (Chile). FEROS covers a spectral interval between 3\,500 and 9\,200 \AA, providing high-resolution spectroscopic data with a resolving power $R=\lambda/\Delta \lambda\sim48\,000$. 

The observed targets are G/K giants with effective temperature ranging from 4\,000 to 5\,500 K and metallicity within the interval $-1.0\lesssim\textrm{[Fe/H]}\lesssim+0.3$, typical of galactic population membership\footnote{Throughout this paper, we adopt the standard spectroscopic notation, ${\rm [e1/e2]}=\log(N_{\rm e1}/N_{\rm e2})_{\star} -\log(N_{\rm e1}/N_{\rm e2})_{\odot}$, and the definition $A({\rm el})=\log\epsilon({\rm el})=\log(N_{\rm el}/N_{\rm H})+12$. We use the [Fe/H] ratio as a proxy of metallicity.} \citep[cf.][]{pereira2011b, decastro2016}. As for the atmospheric parameters, effective temperature ($T_{\rm{eff}}$), surface gravity ($\log g$), microturbulent velocity ($\xi$), and metallicity, we have adopted, for consistency, the same values previously derived in the other papers of this series. These parameters were obtained from the excitation and ionization balances, considering a set of Fe\,{\sc i} and Fe\,{\sc ii} lines. To perform this task, we used the {\sc moog} \citep[][]{sneden1973, sneden2012}{}{}, a {\sc fortran} code that solves the radiative-transfer equation, by assuming a plane–parallel stellar atmosphere and the Local Thermodynamic Equilibrium (LTE) conditions. The 1D plane–parallel model atmospheres of \citet[][]{kurucz1993}{}{} were adopted as inputs, and abundances were normalized to the solar atmosphere values recommended by \citet[][]{grevesse1998}{}{}. They are listed in Table~\ref{tab:ab_data}.

It is worth noting that we have included the star HD~26 as a new target in the present analysis. This star was observed in the same run of the spectroscopic survey. Its high-resolution (FEROS) spectrum was extracted with an exposure time of of 1\,500~s. Since this star was not considered in the other papers of this series, we conducted a full analysis for HD~26, by performing the same approach employed in the previous studies. A full list containing the atmospheric parameters and elemental abundances for HD~26 is provided in the Appendix~\ref{sec:app_hd26}.

\section{Abundances derivation}\label{sec:abundance}

In optical spectra, the [O\,{\sc i}] line at 6\,300~\AA\ and the O\,{\sc i} triplet lines around 7\,774~\AA\ are commonly used as a diagnostic of oxygen abundances. Indeed, based on abundance data derived from the [O\,{\sc i}] line, collected from the literature for a sample of giants and dwarfs, \citet[][]{melendez2002} outlined the Galactic behavior of the [O/Fe] ratios for a metallicity interval ranging from $-3.0$ to $0.0$~dex. Relying on this analysis, we constrained the oxygen abundances for the program stars, assuming here that Ba stars follow the same trend as the normal field stars. That choice rests on theoretical and observational hints, as we list below.

\textit{(i)} Oxygen is mainly a by-product of nucleosynthesis of massive stars \citep[][]{woosley1995, timmes1995, kobayashi2020}. Since Ba stars received the ejecta from ancient TP-AGB stars, they are not expected to show oxygen enhancements. \textit{(ii)} Additionally, we are guided here by observational evidences that Ba stars do not show any appreciable chemical peculiarities for oxygen \citep[see, e.g.,][]{allen2006, karinkuzhi2018b, shejeelammal2020}. \textit{(iii)} Concerning nucleosynthesis of low-mass stars close to solar metallicities, it is worth mentioning that models of \citet[][see their Table~6]{battino2019}, which include convective-boundary mixing at the bottom of the He intershell during the thermal pulse episodes, predict an oxygen enhancement on TP-AGB stars of roughly $\sim50-60$\%, relative to models without extra overshoot. Such a prescription would introduce, at most, a change of $\sim+0.20$~dex in the predicted [O/Fe] ratio. However, after mass transfer, the dilution of the material deposited in the extended envelopes of Ba stars will decrease this value, leading to a negligible contribution from the TP-AGB star to the oxygen abundances observed in Ba stars.

\begin{figure}
    \centering
    \includegraphics[]{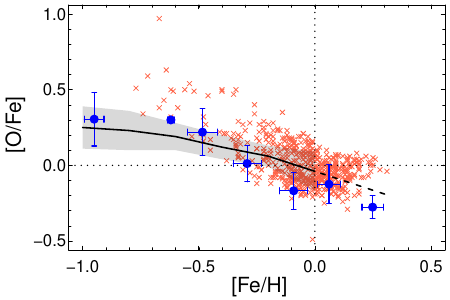}
    \caption{Observed average [O/Fe] ratios (blue dots) for targets selected from the program stars; these objects are representative within each bin of metallicity and cover the entire interval considered in this work, and the error bars represent the standard deviations. The black curve comes from the average [O/Fe] values reported by \citet[][]{melendez2002}, and the shaded area outlines their standard deviations. Data for normal giants (crosses), taken from \citet[][]{luck2007} and \citet[][]{takeda2019}, are also plotted on this diagram.}
    \label{fig:oxygen}
\end{figure}

As a further check of the reasonableness of our assumption, we selected, among the program stars, representative targets within each bin of 0.2~dex in metallicity (see Sec.~\ref{sub_sec:cfe_nfe_feh}), covering the entire range of observation. For these objects, we performed spectral synthesis of the O\,{\sc i} triplet lines. Since the profiles of the two components at 7\,774.2~\AA\ and 7\,775.4~\AA\ present some contribution from CN absorption features, we used the cleanest triplet line at 7\,771.9~\AA\ as oxygen abundance diagnostic. For this transition, we adopted $\log$\textit{gf}$=0.352$, sourced from \citet[][]{ramirez2013}. We then applied to the LTE abundances corrections due to non-LTE (NLTE) effects; these corrections were computed by interpolating the data presented in Table 3 of \citet[][]{takeda2003}. For the 7\,771.9~\AA\ line, these corrections decrease the LTE abundances by $\sim0.10-0.20$~dex.

Figure~\ref{fig:oxygen} shows the average [O/Fe] ratios derived in each metallicity bin (blue dots), together with the data reported by \citet[][black curve]{melendez2002} and data for normal giants (crosses), taken from \citet[][]{luck2007} and \citet[][]{takeda2019}. As one can see, the mean values of the [O/Fe] ratios are in close agreement to the average values of \citeauthor{melendez2002} for normal stars. The blue dots present a mean absolute difference of $0.09\pm0.04$~dex from the black curve. Additionally, inspecting Figure~\ref{fig:oxygen}, it is noteworthy that for $\rm{[Fe/H]}>-0.50$~dex the blue dots present a negative offset from the black curve; conversely, for lower metallicities, they fall above it. In fact, a similar behavior is observed for normal giants with $\rm{[Fe/H]}<-0.50$~dex, which also lie above the average data reported by \citet[][]{melendez2002}.

\begin{figure}
    \centering
    \includegraphics[]{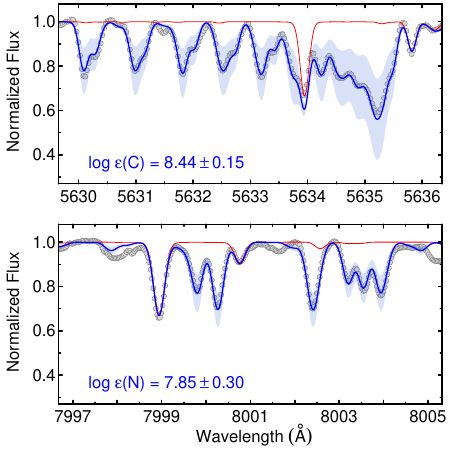}
    \caption{Observed (dots) and synthetic (solid lines) spectra around the spectral region of the C$_{2}$ molecular band at 5\,635~\AA\, (top panel) and CN absorption features at 8\,000~\AA\, (bottom panel) for HD\,107541. The carbon and nitrogen abundances that provide the best fits (blue lines) are indicated in the panels. The shaded blue areas show the effects in changing the carbon and nitrogen abundances around the adopted values. The red lines show spectral synthesis without contribution of the C$_{2}$ and CN molecules.}
    \label{fig:syn_cn}
\end{figure}

Elemental abundances of carbon and nitrogen, as well as the carbon isotopic ratio, were derived by fitting synthetic spectra to the observed absorption features. For this assignment, we run the \textit{synth} drive of {\sc moog}. By inserting an atmospheric model and a line list file containing the laboratory data of the transitions of interest, such as excitation potential and $\log$~\textit{gf} values, {\sc moog} generates a set of LTE spectra, allowing the user a direct comparison to the observed spectrum. The final abundance is the one that provides the best fit between the observed and theoretical spectra. As in the previous papers in this series, we have also adopted here the solar abundances recommended by \citet[][]{grevesse1998}, $\log\epsilon(\rm{C})=8.52$, $\log\epsilon(\rm{N})=7.92$, $\log\epsilon(\rm{O})=8.83$, and $\log\epsilon(\rm{Fe})=7.50$. Additionally, the elemental abundances derived in the other papers were also adopted in the present analysis and kept fixed throughout the spectral synthesis calculations.

For carbon, we used the molecular C$_{2}$ (0,1) band head of the $A^3\Pi_{g} - X^3\Pi_{u}$ system at $\sim5\,635$~\AA. An example of the fitting is illustrated in the top panel of Figure~\ref{fig:syn_cn}, where the theoretical spectra (curves) of the C$_{2}$ molecule computed for different C abundances are superimposed to the observed C$_{2}$ absorption features for HD~107541. This is the most $s$-rich star of the sample. 

Nitrogen abundances were derived from the absorption features within the wavelength range at $\sim7\,995 - 8\,005$~\AA\, due to the $A^2\Pi-X^2\Sigma$ lines of the $^{12}$CN molecule (bottom panel of Figure~\ref{fig:syn_cn}). The carbon isotopic ratios $^{12}$C/$^{13}$C were determined from the $^{13}$CN features at $8\,004 - 8\,005$~\AA. A detailed description of the derivation of the C and N abundances, as well as of $^{12}$C/$^{13}$C ratios, is provided in \citet[][]{drake2008}. The final C, N, and O abundances and $^{12}$C/$^{13}$C ratios adopted here are presented in Table~\ref{tab:ab_data}. Additionally, a full list of the molecular transitions considered in this work is provided in Appendix~\ref{sec:app_linelist}.

\begin{table*}
    \caption{Carbon, nitrogen, and oxygen abundances, along with the carbon isotopic ratios and other selected quantities of interest. The targets are identified in the first column, and their basic atmospheric parameters, $T_{\rm{eff}}$, $\log g$, and [Fe/H], are listed in Columns 2, 3, and 4, respectively. Logarithmic abundances for C, N, O, and C+N are provided in Columns 5, 6, 7, and 8, respectively. C/N and C/O ratios are listed in Columns 9 and 10, respectively. The [X/Fe] ratios are listed in Columns 11, 12, and 13. The carbon isotopic ratios are provided in the last column. This table is fully available in machine-readable format. A small portion is shown here for guidance regarding its form and content.} 
    \centering
    \label{tab:ab_data}
    \begin{tabular}{lccccccccccccc}
    \toprule
                      &  &  &  & \multicolumn{4}{c}{$\log\epsilon(\rm{X})$} &  &  & \multicolumn{3}{c}{[X/Fe]} & \\
                                 \cline{5-8}                                                   \cline{11-13}
        Star        & $T_{\rm{eff}}$ (K) & $\log g$ & [Fe/H] & C & N & O & C+N & C/N & C/O & C & N & O & $^{12}$C/$^{13}$C\\    
    
    \midrule
BD$-08^{\circ}$3194 & 4900  & 3.00 & $-0.10$ &  8.58 & 8.35 & 8.75 & 8.78 & 1.70 & 0.68 & $+0.16$ & $+0.53$ & $+0.02$ &      16  \\
BD$-09^{\circ}$4337 & 4800  & 2.60 & $-0.24$ &  8.47 & 8.08 & 8.65 & 8.62 & 2.45 & 0.66 & $+0.19$ & $+0.40$ & $+0.06$ &       7  \\
BD$-14^{\circ}$2678 & 5200  & 3.10 & $+0.01$ &  8.56 & 8.45 & 8.83 & 8.81 & 1.29 & 0.54 & $+0.03$ & $+0.52$ & $-0.01$ & $\geq32$ \\
CD$-27^{\circ}$2233 & 4700  & 2.40 & $-0.25$ &  8.38 & 8.15 & 8.64 & 8.58 & 1.70 & 0.55 & $+0.11$ & $+0.48$ & $+0.06$ & $\geq28$ \\
CD$-29^{\circ}$8822 & 5100  & 2.80 & $+0.04$ &  8.51 & 8.55 & 8.85 & 8.83 & 0.91 & 0.46 & $-0.05$ & $+0.59$ & $-0.02$ &      19  \\
...                 & ...   & ...  & ...     &  ...  & ...  & ...  & ...  & ...  & ...  & ...     & ...     & ...     & ...      \\
    \bottomrule
    \end{tabular}
\end{table*}

\subsection{Uncertainty estimates}

To assess uncertainties, we organized the program stars into three different ranges of effective temperatures, similarly to the previous studies in this series. For each group, we selected a representative star: BD$-14^{\circ}2678$, HD~119185, and HD~130255. Then, the uncertainty estimates performed for these three targets were applied to all stars belonging to the respective group. In the task of evaluating abundance uncertainties, we have taken into account uncertainties associated with the atmospheric parameters and the CNO abundances themselves, since the abundances of these elements are interdependent on each other. A similar procedure was carried out, for example, in \cite{roriz2023}. 

In detail, we varied each of the atmospheric parameters, keeping the others fixed, and computed the respective change introduced in the C and N abundances. Furthermore, to consider the interdependence of the C and N abundances, we evaluated the changes introduced in them as a consequence of shifts of $+0.20$~dex in the C, N, and O abundances. The results are presented in Table~\ref{tab:erros} for each template star. By adding the square of the individual uncertainties and extracting the square root, we evaluated the final uncertainty in $\log\epsilon(\rm{X})$, shown in the last column of Table~\ref{tab:erros}. As the oxygen abundances were derived from the parameterization of \citet[][]{melendez2002}, uncertainties in $\log\epsilon(\rm{O})$ were not estimated.

For the template stars, BD$-14^{\circ}2678$, HD~119185, and HD~130255, just lower limits for the carbon isotopic ratios were derived. To estimate the uncertainties in the $^{12}$C/$^{13}$C ratios, we selected the stars HD~91208 (5\,100~K), HD~116869 (4\,800~K), and HD~123396 (4\,600~K), which have atmospheric parameters similar to BD$-14^{\circ}2678$, HD~119185, and HD~130255, respectively. We then evaluated the changes introduced in the $^{12}$C/$^{13}$C ratios as a response to a shift of $+100$~K in effective temperatures, which yielded $\Delta^{12}\rm{C}/^{13}\rm{C}=-6$, $-3$, and $-3$ for HD~91208, HD~116869, and HD~123396, respectively.

\begin{table*}
	\caption{Abundance uncertainties for the template stars BD$-14^{\circ}2678$, HD~119185, and HD~130255. Columns 2 to 7 show the variations introduced in the abundances as a consequence of changes in $T_{\rm eff}$, $\log g$, [Fe/H], $\log\epsilon(\rm{C})$, $\log\epsilon(\rm{N})$, and $\log\epsilon(\rm{O})$, respectively. By combining quadratically the terms from the 2$^{nd}$ to 7$^{th}$ column, we estimate the total uncertainty, listed in Column 8.}
    \centering
	\label{tab:erros}
	\begin{tabular}{lcccccc|c}
        \toprule
        \multicolumn{8}{c}{BD$-14^{\circ}2678$} \\
		\bottomrule
		 & $\Delta T_{\textrm{eff}}$ & $\Delta \log{g}$ & $\Delta$[Fe/H] & $\Delta\log\epsilon(\rm{C})$ & $\Delta\log\epsilon(\rm{N})$ & $\Delta\log\epsilon(\rm{O})$  \\
            &  ($+100$~K)  &  ($+0.20$)  &  ($+0.10$)  &  ($+0.20$)  &  ($+0.20$)  &  ($+0.20$)  &  $\sqrt{\Sigma \sigma^{2}}$\\
		\midrule

        $\Delta\log\epsilon(\rm{C})$   &  $+$0.05  &     0.00  & $+$0.02  & --       &  0.00   &  $+$0.05  & 0.07  \\
        $\Delta\log\epsilon(\rm{N})$   &  $+$0.15  &     0.00  & $+$0.08  & $-$0.22  &  --     &  $+$0.10  & 0.30 \\

        \toprule
        \multicolumn{8}{c}{HD~119185} \\
		\bottomrule
            &  ($+100$~K)  &  ($+0.20$)  &  ($+0.10$)  &  ($+0.20$)  &  ($+0.20$)  &  ($+0.20$)  &  $\sqrt{\Sigma \sigma^{2}}$\\
		\midrule

        $\Delta\log\epsilon(\rm{C})$   &  $+$0.03  &  $+$0.03  & $+$0.04  & --       &  0.00  &  $+$0.06  & 0.08  \\
        $\Delta\log\epsilon(\rm{N})$   &  $+$0.12  &  $+$0.05  & $+$0.09  & $-$0.20  &  --    &  $+$0.10  & 0.27 \\

        \toprule
        \multicolumn{8}{c}{HD~130255} \\
		\bottomrule
            &   ($+90$~K)  &  ($+0.20$)  &  ($+0.10$)  &  ($+0.20$)  &  ($+0.20$)  &  ($+0.20$)  &  $\sqrt{\Sigma \sigma^{2}}$ \\
        \midrule

        $\Delta\log\epsilon(\rm{C})$   &     0.00  &  $+$0.05  & $+$0.05  & --       &  0.00  &  $+$0.15  & 0.17  \\
        $\Delta\log\epsilon(\rm{N})$   &  $+$0.10  &  $+$0.05  & $+$0.05  & $-$0.20  &  --    &  $+$0.30  & 0.38  \\

        \bottomrule
	\end{tabular}
\end{table*}

\section{Abundance analysis}\label{sec:analysis}

\subsection{[C/Fe] and [N/Fe] ratios versus [Fe/H]}\label{sub_sec:cfe_nfe_feh}

\begin{figure}
    \centering
    \includegraphics[]{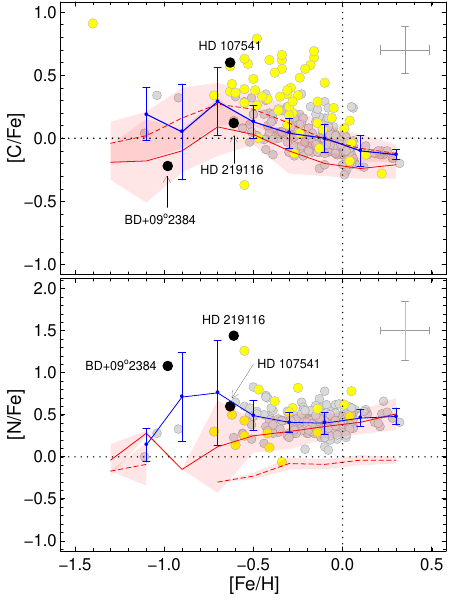}
    \caption{Carbon (\textit{top}) and nitrogen (\textit{bottom}) abundance ratios to Fe derived for the program stars (grey dots) as a function of metallicity. Typical error bars are shown in top right side of the panels. The lines outline
    %the general behavior
    the average [C/Fe] and [N/Fe] ratios observed in our sample of Ba giants (blue solid line), normal field giants (red solid line) and normal field dwarfs (red dashed line). The error bars in the blue lines represent the standard deviations of the mean values observed in each metallicity bin. The shaded red areas reflect the standard deviations of the literature data for dwarf and giant normal stars. Specifically for nitrogen abundances in normal dwarfs with $[\rm{Fe/H]}>-0.70$~dex, we have considered the homogeneous data of \citet[][]{takeda2023}. The stars BD$+09^{\circ}2384$, HD~107541, and HD~219116 are identified as black dots (see text). For comparison purposes, abundance data observed in Ba dwarfs (yellow dots) are also plotted in this figure; data were taken from \citet[][and references therein]{roriz2024_dba}.}
    \label{fig:cno}
\end{figure}

In Figure~\ref{fig:cno}, the [C/Fe] and [N/Fe] ratios derived for the program stars are plotted (gray dots) as a function of metallicity. The [C/Fe] ratios increase from $\sim -0.30$ to $+0.60$~dex with decreasing metallicity. The [N/Fe] ratio is instead relatively flat, with [N/Fe] around $+0.50$~dex, although a slight increase in [N/Fe] appears below $\rm{[Fe/H]}\lesssim-0.30$~dex. These features become even more evident when we average the observational data within bins of $0.20$~dex in metallicity; this is shown by the blue curves drawn in both panels of Figure~\ref{fig:cno}. For the calculated means, their standard deviations are shown as error bars in the curves. The carbon and nitrogen abundances reported in the literature for Ba giant stars align with the bulk of our observations \citep[cf.][]{barbuy1992, allen2006, karinkuzhi2018b, shejeelammal2020}. 

The red curves in Figure~\ref{fig:cno} outline the general trends observed for normal giants (solid line) and normal dwarfs (dashed line) within the metallicity range of the program stars. These profiles were derived from the carbon and nitrogen abundance data reported in the literature by different studies. Similarly, we grouped the literature data into bins of $0.20$~dex in metallicity and evaluated the mean value in each bin. For giants, data were collected from \citet[][]{barbuy1988}, \citet[][]{sneden1991}, \citet[][]{carretta2000}, \citet[][]{gratton2000}, \citet[][]{simmerer2004}, \citet[][]{spite2005}, \citet[][]{mishenina2006}, \citet[][]{luck2007}, and \citet[][]{takeda2019}, which comprises a sample of $\sim870$ normal giants, thus reflecting the local behavior of the Galactic field. For dwarfs, we sampled $\sim960$ targets from data reported by \citet[][]{reddy2003}, \citet[][]{reddy2006}, \citet[][]{luck2006}, and \citet[][]{takeda2023}. In order to mitigate systematics affecting data from different analyses, the abundance ratios of normal giants and normal dwarfs were scaled to the solar values from \citet[][]{grevesse1998}, as adopted in this work.

When comparing the carbon trends observed for Ba giants, normal giants, and normal dwarfs in the top panel of Figure~\ref{fig:cno}, all the populations exhibit a similar behavior, with carbon abundances increasing for lower metallicities and reaching a maximum at $\rm{[Fe/H]}\sim-0.70$~dex. Such a behavior is an effect due to galactic chemical evolution \citep[e.g.][]{romano2022}. The difference between the field dwarfs and the field giants is due to the First Dredge-Up (FDU) during the red giant branch (RGB), which decreases the C abundance by bringing to the stellar atmospheres the by-products of the CN-cycle \citep[][and references therein]{iben1967, karakas2014}. Therefore, normal giants are expected to be C-depleted relatively to their dwarf counterparts, which is widely supported by observations. Ba stars, instead, show [C/Fe] ratios systematically larger than the values found in normal giants, with an average shift $\langle\rm{[C/Fe]_{gBa}}-\rm{[C/Fe]_{g}}\rangle=+0.18\pm0.05$~dex in relation to them. Furthermore, as illustrated in Figure~\ref{fig:cno} and indicated by the low standard deviation of this average value, that offset is approximately constant within the metallicity range covered by the program stars. This enhancement observed in Ba stars is expected by pollution from an TP-AGB companion as C is produced during partial He burning in the interiors of TP-AGB stars. Therefore, the [C/Fe] ratios in the program stars being above the levels observed in normal giants is consistent with the mass transfer scenario.

Regarding the nitrogen abundances, the bottom panel of Figure~\ref{fig:cno} shows that Ba giants, normal giants, and normal dwarfs exhibit a similar flat behavior. As in the case of C, the difference between the field giants and dwarfs is due to the FDU. The behavior of the field giants and Ba giants is similar for $\rm{[Fe/H]}\gtrsim-0.30$~dex. For lower metallicities, the mean [N/Fe] ratios observed in Ba stars indicate an increase, in contrast with normal giants, which show lower [N/Fe] ratios. This behavior may be due to low statistics, as the number of targets is limited in that metallicity range, or be the consequence of the activation of extra-mixing processes below the base of the convective envelope of the lowest metallicity TP-AGB parent stars. This process, also known as Cool Bottom Processing \citep[CBP; e.g.,][]{nollett2003}, would need to have been present in the TP-AGB phase of the companion, as there is no indication of its occurrence in the field giants. The N increase is not expected to be due to Hot Bottom Burning \citep[HBB; i.e., proton captures at the base of the convective envelope, e.g.][]{sackmann1992} in a massive ($\gtrsim4~M_{\odot}$) TP-AGB companion, as this would produce positive [Rb/Zr]. As reported in Paper~II, instead, we found $\rm{[Rb/Zr]}<0$ for the whole sampled stars, which is a proxy of the companion mass and indicates low-mass TP-AGB stars \citep[see][]{vanraai2012}.

Three special objects are identified as black dots in Figure~\ref{fig:cno} and deserve special attention. In our sample, HD~107541 is the most C-rich star, with $\rm{[C/Fe]}=+0.60$~dex; it is also the most $s$-rich program star, as will be discussed later. The nitrogen abundance of HD~107541 is typical for its metallicity. BD$+09^{\circ}$2384 and HD~219116, instead, are the most N-rich program stars, with $\rm{[N/Fe]}=+1.08$ and $+1.44$~dex, respectively, whereas their carbon contents follow the data cloud. Such high N abundances were also reported in Ba stars \citep[cf.][]{karinkuzhi2018b, shejeelammal2020}, as well as in some Carbon-Enhanced Metal-Poor (CEMP) stars \citep[e.g.][]{karinkuzhi2021}. As discussed above, this could indicate the high mass nature of their former TP-AGB companions, although their negative [Rb/Zr] ratios favor the extra-mixing scenario. 

\subsection{C and N abundances in light of FDU and mass transfer hypothesis}

In Figure~\ref{fig:cno} we also show observational data for the Ba dwarfs (yellow dots), which are thought to be the progenitors of the classical Ba giants \citep[][]{tomkin1989, luck1991, north1994}. Although the two populations share many similarities, the current sample of Ba dwarfs identified from high-resolution spectroscopy amounts only 71 targets, therefore significantly smaller than the sample of their giant counterparts \citep[cf.][]{roriz2024_dba}. 

As depicted in Figure~\ref{fig:cno}, Ba dwarfs show [C/Fe] ratios systematically larger than the trends observed for both Ba giants and normal dwarfs, in addition to a large spread. This is expected as these stars did not go through the FDU yet and therefore represent the unaltered signature of the material accreted from the TP-AGB companion. In terms of C, they are a more direct indication of the AGB nucleosynthesis. By comparing the C abundances observed in the Ba giants and Ba dwarfs, we estimate an average shift $\langle\rm{[C/Fe]_{gBa}}-\rm{[C/Fe]_{dBa}}\rangle=-0.19\pm 0.13$~dex in the trend of Ba giants due to the FDU. This is an indication that although the FDU lowered the C abundances in the Ba giant stars, it did not completely \textit{erase} the C-rich signature of the material previously transferred by the former TP-AGB companions, as the Ba giants still have higher C than the field giants. 

Although detailed models are missing for giant stars with accreted AGB material, this observational feature can nevertheless be evaluated with simple dilution calculations. If we consider a 1~M$_{\odot}$ normal giant star with an envelope of $\sim0.8$~M$_{\odot}$, the effect of the FDU observed in the sample (i.e., a decrease of $\sim0.2$~dex in C from the normal stars) can be obtained by mixing $\sim0.5$~M$_{\odot}$ of original solar $^{12}$C abundance with a $\sim0.3$~M$_{\odot}$ of material depleted in $^{12}$C \citep[located in the inner part, close to the core; see Figure 5 of][]{karakas2014}. As a second step, it is possible to obtain the observed difference in the C abundance between the Ba giants and the normal giants by mimicking the effect of accreted AGB C-rich material (typically 4 times the solar value) by mixing a third component with accreted mass of 0.1~M$_{\odot}$. This value appears realist.

For $\rm{[Fe/H]}\gtrsim0.0$~dex, the FDU effect between the Ba dwarfs and the Ba giants is less pronounced than for the lower metallicities. This is consistent with TP-AGB models that usually predict higher C abundances in stars as the metallicity decreases \citep[e.g.][]{karakas2016}. The [N/Fe] ratios in some Ba dwarfs show the same typical values observed in the Ba giants, and larger than those observed in normal dwarfs. This is probably because the companion TP-AGB stars underwent the FDU. A clear indication of extra-mixing on the former TP-AGB stars companions of Ba dwarfs, however, does not appear, except for HD~94518 (the yellow dot close to HD~219116 in Figure~\ref{fig:cno}) with the clearly highest [N/Fe], reported by \citet[][]{shejeelammal2020}.

\begin{figure}
    \centering
    \includegraphics[]{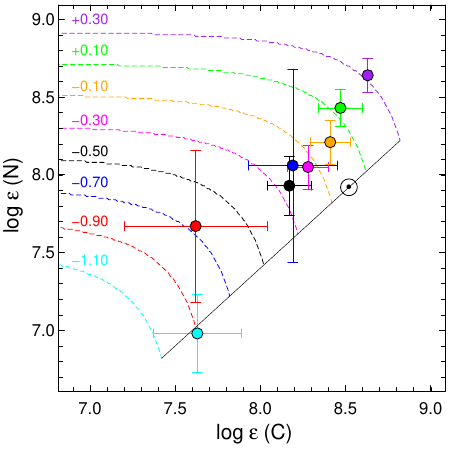}
    \caption{Expected evolution (dashed curves) of the photospheric abundances of carbon and nitrogen, as the star leaves the main-sequence and becomes a red giant. For different metallicity regimes, identified by colors in this figure and labeled in the curves, these tracks take into account only mixing due to FDU, which keep constant the C+N abundances. In this plot, the dots represent the mean C and N abundances observed in our targets, computed in bins of $0.20$~dex in metallicity, centered on the metallicity regimes labeled in the curves. Error bars represent the standard deviation of the mean. The black straight line represents the solar C and N abundances scaled to metallicity. In general, the observational data present an offset from the track where they were expected to lie (see text for more explanations). The solar values are marked by the symbol $\odot$.}
    \label{fig:cn_smith}
\end{figure}

In the framework of an isolated star, although FDU changes its initial C and N photospheric abundances when the star leaves the main-sequence and ascends to RGB, the $\log\epsilon(\rm{C+N})$ quantity, however, remains constant. With this in mind, we drawn in Figure~\ref{fig:cn_smith} tracks that mimic the FDU effect on the C and N abundances for different regimes of metallicities \citep[i.e., changing the C and N abundances, but keeping constant the sum C+N; see][]{smith2002, smith2013}. On the same plane, the average values of $\log\epsilon(\rm{C})$ and $\log\epsilon(\rm{N})$ observed in each bin of metallicity are plotted.

In general, the binned data show a shift from the curve where they were expected to lie. Such a feature is observed to a lesser extent in stars with $\rm{[Fe/H]}\gtrsim0.0$, whereas larger deviations are observed for lower metallicity regimes. As discussed above, this is due to the enhanced C abundance. For stars with $\rm{[Fe/H]}\sim-0.70$ (regime indicated in blue in Figure~\ref{fig:cn_smith}), the mismatch is the most evident. This is mainly due to the increase observed in N abundances for stars within that metallicity regime. As a consequence, these objects are shifted to more metal-rich regions in this diagram. Even without considering the most N-rich program stars, BD$+09^{\circ}2384$ and HD~219116, in the computation of the mean C and N abundances, we were unable to reconcile the predictions with the observations.

\subsection{CNO abundances and carbon isotopic ratio}

\begin{figure*}
    \centering
    \includegraphics[]{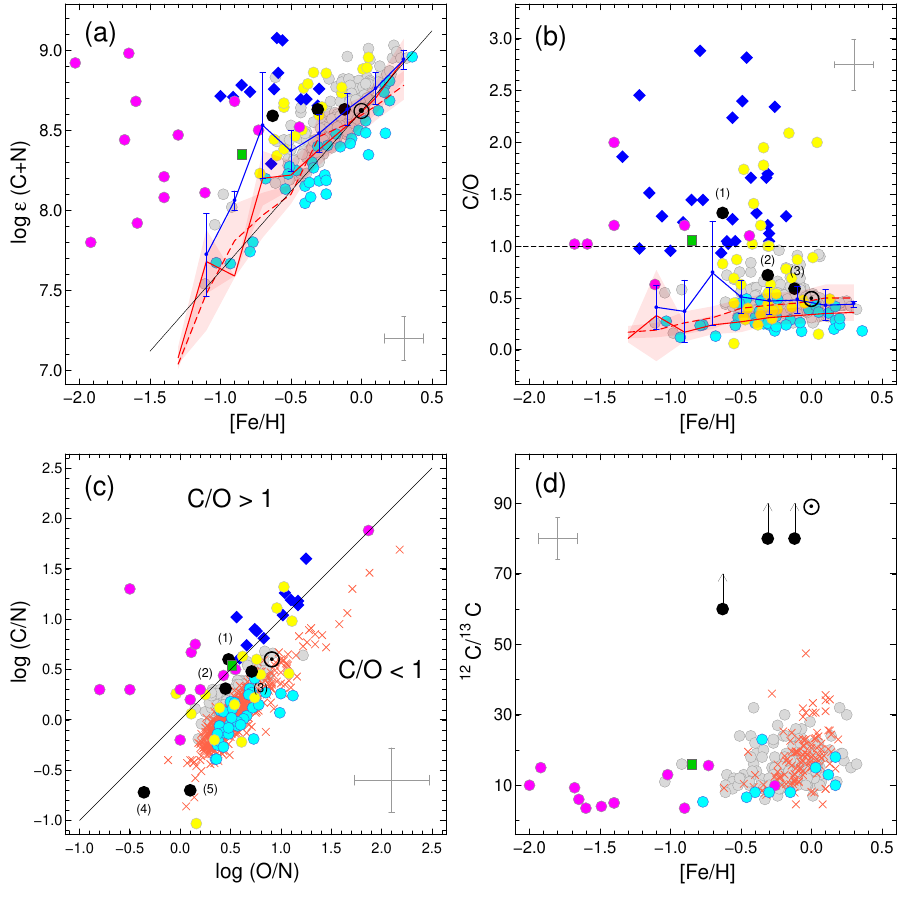}
    \caption{\textit{Panel (a)}: $\log\epsilon(\rm{C+N})$ versus [Fe/H]; \textit{panel (b)}: C/O ratios versus [Fe/H]; \textit{panel (c)}: $\log{(\rm{C/N})}$ versus $\log{(\rm{O/N})}$; \textit{panel (d)}: $^{12}$C/$^{13}$C ratios versus [Fe/H]. The respective solar values are marked by the $\odot$ symbol. As in Figure~\ref{fig:cno}, grey dots are data for the Ba giants of this study and yellow dots are data for Ba dwarfs. The curves outline the average trends observed for the program stars (blue solid line), normal giants (red solid line), and normal dwarfs (red dashed line). The shaded red areas reflect the standard deviations of the literature data for dwarf and giant normal stars. The stars HD~107541 (1), HD~66291 (2), HD~39778 (3), HD~219116 (4), and BD$+09^{\circ}$~2384 (5), mentioned in the text, are plotted as black dots. HD~26 is shown as a green square. In panel (a), the black straight line represents the solar scaled C+N abundance. In panel (d), data for which just lower limits were estimated are not shown, except for except for HD~107541, HD~66291, and HD~39778. In this figure, crosses represent data for normal giants, taken from \citet[][]{luck2007} and \citet[][just abundances labeled as `reliable']{takeda2019}. Additionally, data are shown for symbiotic stars \citep[cyan dots;][]{galan2016, galan2017}, classical CH stars \citep[magenta dots;][]{vanture1992b, pereira2009, pereira2012b, goswami2016, purandardas2019}, and post-AGB stars \citep[blue diamonds;][]{vanwinckel2000, reyniers2004, desmedt2012, desmedt2015, desmedt2016, vanaarle2013}.}
    \label{fig:ratios}
\end{figure*}

In panel (a) of Figure~\ref{fig:ratios} we plot the composed C+N abundances derived for the program stars versus metallicity. We then compare the trend observed in our data set with the trends observed for normal giants and normal dwarfs. In addition to the data for Ba dwarf stars (yellow dots), we also included in this figure observational data for classical CH stars (magenta dots), post-AGB stars (blue diamonds), and S-type symbiotic stars (cyan dots), which represent another population of binary stars (i.e., a M-type star and a current white dwarf). The S-type symbiotic stars may be seen as a possible counter example that not all related binary systems will show any signature of AGB nucleosynthesis \citep[see, e.g.][]{pereira2017}, showing that the mass-transfer phenomenon still lacks a complete understanding \citep[see also][]{jorissen2003}. Data for the so-called yellow symbiotics, which show $s$-process pollution, resembling the Ba stars \citep[][]{jorissen2005}, are not plotted in Figure~\ref{fig:ratios}. The solar scaled C+N abundances are shown as a black straight line in this diagram. Once again, we see that Ba stars present $\log\epsilon(\rm{C+N})$ systematically greater than normal giants (as in Figure~\ref{fig:cno}). Since Ba stars have their envelopes contaminated by carbon, the $\log\epsilon(\rm{C+N})$ quantity is no longer expected to remain constant. Additionally, the composed C+N abundances observed in Ba giants are typically slightly smaller than the values reported in the literature for their unevolved counterparts, the Ba dwarfs. This is because the Ba giants diluted the original C enrichment in their convective envelope, whereas for Ba dwarfs the dilution depends on the depth of the surface convection of the stars, therefore on their mass. Although data for CH stars present a large spread, they show $\log\epsilon(\rm{C+N})$ at the same level of Ba giant stars. The symbiotic stars, in turn, closely follow the solar scaled C+N abundance. 

In panel (b) of Figure~\ref{fig:ratios}, we show the C/O ratios observed in our Ba giants as a function of metallicity. The average trends observed for the program stars, normal giants, and normal dwarfs are also shown. The classical CH stars are known as the metal-poor counterparts of the Ba giants, and generally show $\rm{C/O}>1$, as well as the $s$-rich CEMP stars (CEMP-$s$). Ba stars, on the other hand, show $\rm{C/O}<1$. Indeed, this feature is observed in all the program stars, except for HD~107541, with $\rm{C/O}=1.32$ ($\rm{[Fe/H]}=-0.63$~dex), which allows us to classify it as a CH star. Interestingly, there are also some Ba dwarfs with $\rm{C/O}>1$; however, except for a few outliers, Ba dwarfs behave similarly to the Ba giants, in agreement with the idea that the latter descend from the former. Post-AGB stars, which are expected to be the progeny of the TP-AGB stars that pollute the Ba and CH stars, show $\rm{C/O}\gtrsim1$. We also observe that symbiotic stars show C/O ratios lower than the values observed in Ba giants, closely resembling the trend observed for normal giants.

The magenta dot below the line $\rm{C/O}=1$ in panel (b) of Figure~\ref{fig:ratios} corresponds to the observational data reported by \citet[][]{goswami2016} for HD~26 ($\rm{[Fe/H]}=-1.11$~dex; $\rm{[C/Fe]}=+0.31$~dex; $\rm{C/O}=0.63$). This star is sometimes referred to as a prototypical CH star. Other authors have found $\rm{C/O}$ ratios very close to unity for HD~26 \citep[cf.][]{vanture1992b, masseron2010, karinkuzhi2021}. The last two analyses, in particular, qualify HD~26 as a CEMP-$s$ star (these data are not plotted in Figure~\ref{fig:ratios}). In fact, from a chemical perspective, the criteria for classifying a star as CH or CEMP-$s$ may vary slightly from author to author. As mentioned in Section~\ref{sec:sample}, HD~26 has been included in our analysis and subjected to a full analysis (see Appendix~\ref{sec:app_hd26}). We found for HD~26 a moderated C-enhancement ($\rm{[C/Fe]}=+0.57$~dex) and $\rm{C/O}=1.06$. These values are greater than those reported by \citet{goswami2016} and are plotted as a green square in Figure~\ref{fig:ratios}. Additionally, the well known $s$-rich nature of HD~26 is corroborated here, with an index $\rm{[\textit{s}/Fe]}=+1.53$~dex. Although slightly more metal-poor ($\rm{[Fe/H]}=-0.85$~dex), HD~26 closely resembles HD~107541, except for the carbon isotopic ratio, which is lower in HD~26 (see bottom panel of Figure~\ref{fig:ratios}). In panel (c) of Figure~\ref{fig:ratios}, the $\log{(\rm{C/N})}$ versus $\log{(\rm{O/N})}$ plane explicit the carbon rich ($\rm{C/O}>1$) and oxygen rich ($\rm{C/O}<1$) regions. Special objects are also identified in this diagram as black dots.

The panel (d) of Figure~\ref{fig:ratios} shows the carbon isotopic ratios observed in our sample as a function of metallicity, together with data for normal giants. Previous studies have shown that Ba stars have $^{12}$C/$^{13}$C$<20$ \citep[see][]{barbuy1992, karinkuzhi2018b, shejeelammal2020}. Such a feature indicates that the effect of the CN-cycle within the current Ba star (brought to its surface via FDU) dominates over the effect of the transfer of $^{12}$C from its companion. In fact, of the 145 program stars for which $^{12}$C/$^{13}$C could be determined (i.e., neglecting those for which only lower limits were estimated), $83\%$ exhibit $^{12}$C/$^{13}$C$\leq20$ and $17\%$ show $20<^{12}$C/$^{13}$C$\leq40$. Three stars present $^{12}$C/$^{13}$C$\gtrsim60$; they are HD~66291, HD~39778, and HD~107541. Their nitrogen abundances correspond to $\rm{[N/Fe]}=+0.54$, $+0.23$, and $+0.60$~dex, respectively, which are relatively close to the means evaluated in their metallicity ranges.

\subsection{Observational trends with the s-process}

\begin{figure}
    \centering
    \includegraphics[]{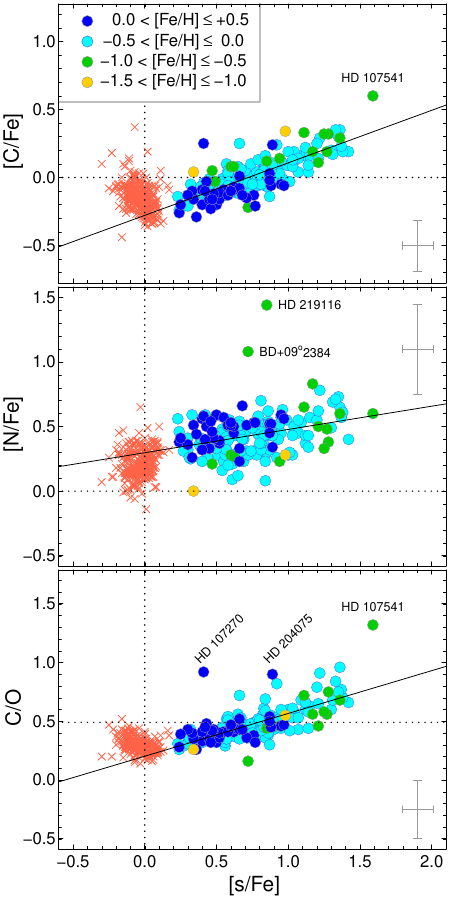}
    \caption{[C/Fe], [N/Fe], and C/O ratios (\textit{top}, \textit{middle}, and \textit{bottom} panels, respectively) as a function of the average $s$-process abundance, [$s$/Fe]. The observational data set (dots) was grouped within different metallicity ranges, as identified by the legend shown in the top panel. Some outlier stars are also identified. Typical error bars of these quantities are also drawn. The black straight lines are linear fits of the data. For comparison, data for normal field giants (crosses), taken from \citet[][]{luck2007}, are plotted in these panels.}
    \label{fig:sfe_cfe}
\end{figure}

In addition to being carbon producers, low-mass ($\lesssim~3.0~\rm{M}_{\odot}$) TP-AGB stars are the astrophysical sites of the main component of the $s$-process \citep[][]{gallino1998}. From an observational point of view, the abundance profiles of Ba stars evidence the low-mass nature of their polluters \citep[e.g.][among others]{cseh2018, karinkuzhi2018b, shejeelammal2020, roriz2021a}. Therefore, a correlation between [C/Fe] and $s$-element abundances is expected. Keeping this in mind, we consider the relationship between the [C/Fe] ratios and the average $s$-process abundance ratios, [$s$/Fe]. The [$s$/Fe] index is calculated as the mean of the [X/Fe] ratios for the elements Sr, Y, Zr, La, Ce, and Nd. This correlation is depicted in the top panel of Figure~\ref{fig:sfe_cfe}, which shows that [C/Fe] is well correlated with [$s$/Fe]. Indeed, a statistical evaluation of that correlation yields a high Pearson coefficient ($\rho_{\rm{P}}=+0.78$). A linear fitting of the data provides $\rm{[C/Fe]}=(+0.39\pm0.02)\times\rm{[\textit{s}/Fe]}-(0.28\pm0.02)$. When we plot [N/Fe] versus [$s$/Fe] (middle panel of Figure~\ref{fig:sfe_cfe}), the data show a larger spread, with a less evident trend ($\rho_{\rm{P}}=+0.32$). A linear fit provides $\rm{[N/Fe]}=(+0.18\pm0.04)\times\rm{[\textit{s}/Fe]}+(0.30\pm0.03)$. For completeness, data for normal field giants (crosses) are also plotted in Figure~\ref{fig:sfe_cfe}.

In light of the stellar evolution models, after a number of Third Dredge-Up (TDU) episodes, which depend on the stellar mass and metallicity and take place after each thermal-pulses, a TP-AGB star can eventually become a carbon star ($\rm{C/O}>1$). TDU is the mechanism by which C and $s$-elements are brought to the external layers of the TP-AGB surface. Therefore, we also expect to find a correlation between C/O and [$s$/Fe] in Ba stars. In fact, this is observed in the program stars, as demonstrated in the bottom panel of Figure~\ref{fig:sfe_cfe}. A linear fit provides $\rm{C/O}=(+0.36\pm0.03)\times\rm{[\textit{s}/Fe]}+(0.20\pm0.02)$, with $\rho_{\rm{P}}=0.70$. 

\subsection{[C/Fe] and [N/Fe] ratios versus [Na/Fe]}\label{sub:na}

\begin{figure}
    \centering
    \includegraphics[]{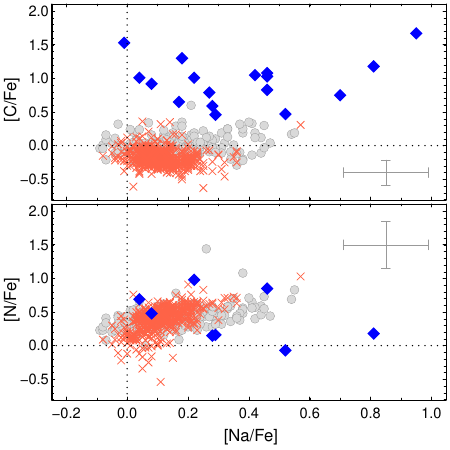}
    \caption{Carbon (\textit{top panel}) and nitrogen (\textit{bottom panel}) abundance ratios to Fe versus sodium abundances observed in Ba giants (grey dots); typical error bars are shown. Data for normal giants \citep[crosses; taken from][]{luck2007, takeda2019} and post-AGB stars (blue diamonds; see references in the caption of Figure~\ref{fig:ratios}) were added in these plots.}
    \label{fig:nna}
\end{figure}

In this section, we explore the novel abundance results together with the sodium abundances previously reported in Paper~I. In Figure~\ref{fig:nna}, we plot the [C/Fe] and [N/Fe] ratios observed in the program stars versus the [Na/Fe] ratios, along with data for normal red giants and post-AGB stars. In general, we observe a similar behavior in the Ba stars and normal giants, in that carbon abundances in Ba stars exhibit a very weak correlation with sodium abundances ($\rho_{\rm{P}}=+0.17$), whereas nitrogen abundances present a moderate correlation ($\rho_{\rm{P}}=+0.47$) with sodium abundances. Nevertheless, in addition to an overall offset observed in the [C/Fe] ratios of Ba giants relative to the normal red giants, attributable to the material accreted, $\sim17\%$ of the program stars show Na enhancement features, with $\rm{[Na/Fe]}>+0.30$~dex, as noticed in Paper~I. Other studies have also reported Na enhancements in the envelopes of Ba stars \citep[e.g.][among others]{allen2006, karinkuzhi2018b, shejeelammal2020, roriz2023}. Notably, as depicted in Figure~\ref{fig:nna}, some post-AGB stars also present Na enhancements \citep[see also][]{pereira2012a}.

Sodium is mainly produced by hydrostatic carbon burning of massive stars ($10-40~M_{\odot}$) through the $\rm{^{12}C(^{12}C,p)^{23}Na}$ reaction \citep[][]{woosley1995}. Additionally, this odd-Z nuclide can be synthesized by the NeNa chain, involving proton captures, via the $^{22}\textrm{Ne}(\textrm{p},\gamma)^{23}\textrm{Na}$ reaction, which takes place in H-core burning of stars with masses $M\gtrsim1.5~\rm{M}_{\odot}$. Then, FDU brings the by-products to the stellar surface when the star becomes a giant \citep{karakas2014}. Therefore, sodium, along with carbon and nitrogen, is an element sensitive to internal evolutionary processes of the stars. In this way, the correlation we found between [N/Fe] and [Na/Fe] in Ba stars is an observational indicative of the NeNa cycle operation in the programme stars.

TP-AGB stars also provide a conducive environment for the sodium production. During the H-shell burning periods (interpulse), $^{23}\rm{Na}$ is expected to be produced at the end of each TDU inside the partial mixing zone that lead to the formation of the $^{13}$C pocket, via proton captures on the abundant $^{22}$Ne \citep[cf.][]{goriely2000, cristallo2009}. Therefore, the sodium content observed in Ba stars may be considered, in principle, as an outcome of both internal nucleosynthesis and external pollution. Close to solar metallicity, $s$-process nucleosynthesis models predict no more than $\rm{[Na/Fe]}\sim+0.20$~dex for low-mass AGB stars \citep[see also][]{karakas2016}. Additionally, the $s$-processed material transferred to the observed Ba star is further diluted in its extended envelope. By fitting predicted abundance profiles to those observed in Ba stars, \citet[][]{cseh2022} reported $\delta\lesssim0.3$, implying in $\rm{[Na/Fe]_{dil}}<0.07$~dex. Therefore, we do not expect a significant contribution from AGB nucleosynthesis in the Na content observed in Ba stars close to solar metallicites. In this metallicity regime, sodium in Ba stars mostly reflects the operation of the FDU. The absence, and the presence, of a correlation with C, and N, respectively, confirms this conclusion.

From the observational point of view, \citet[][]{boyarchuk2001} noticed that normal red giants of lower $\log~g$ values show higher [Na/Fe] ratios. On the other side, such a behavior is not observed in the samples of \citet[][]{mishenina2006} and \citet[][]{luck2007}, whereas it is apparent in the giants of \citet[][]{takeda2008}. In Paper~I, it was noticed an anticorrelation between [Na/Fe] and $\log~g$ for our sample (see its Figure 20). Aligned to this, although with a slightly larger spread, it is worth noting that data of \citet[][]{allen2006}, \citet[][]{karinkuzhi2018b}, and \citet[][]{shejeelammal2020} claim the same behavior. Additionally, taking into account the data gathered by \citet[][]{roriz2024_dba} for Ba dwarfs, we notice that they follow the same trend derived from the giant stars, when the linear fit is extrapolated for larger $\log g$ values. This means that the Na content observed in our targets has a contribution of internal nucleosynthesis of the Ba star itself, whose by-products are brought to the surface via FDU, as the star ascends the giant branch. On the other hand, as reported in Paper~I, our data also reveal that the [Na/Fe] ratios increase with decreasing metallicities. This is also observed in other studies \citep[cf.][]{allen2006, karinkuzhi2018b, shejeelammal2020}, and may be attributable to contributions from the parent TP-AGB stars, since some AGB models yield larger amounts of Na for lower metallicities. Therefore, Na enhancements from mass transfer for Ba stars are expected in these regimes.

\section{Comparison to nucleosynthesis models}\label{sec:models}

It is not possible to compare the C, N, and carbon isotopic ratio reported here directly to the AGB models. This is because the abundances accreted from the AGB companion are subsequently affected by the FDU and the potential extra-mixing processes that change the C and N abundances and carbon isotopic ratio during the red giant phase. Nevertheless, it is interesting to compare AGB models to the objects with C abundances and C/O ratios significantly higher (beyond the error bar) than the other stars in the same metallicity range. Considering the bottom panel of Figure~\ref{fig:sfe_cfe}, these are: HD~107541 (the green dot with the highest C/O = 1.32), HD~107270 and HD~204075 (the two blue dots with C/O just below 1), and CPD$-64^{\circ}4333$ and HD~24035 (the two cyan dots also with C/O just below 1). 

As in the other papers of this series, we conduct our discussion in the light of the Monash \citep[][]{karakas2016, karakas2018} and FRUITY \citep[][]{cristallo2011} nucleosynthesis models, which cover a wide range of mass (1.0 - 8.0~M$_{\odot}$) and metallicity ($-1.20\lesssim\rm{[Fe/H]}\lesssim+0.30$). These models are based on different stellar evolution codes and adopt different physics and nuclear input data. However, the key differences between them lie in their prescriptions to form the $^{13}$C-pocket and to compute the detailed $s$-process nucleosynthesis. In the Monash models, the $^{13}$C pocket forms from a parametric approach, and a post-processing code yields the nucleosynthesis of heavy elements. In the FRUITY models, the $^{13}$C pocket is self-consistently formed and the full nucleosynthesis computation is coupled to the stellar evolution code.

The comparison between the AGB model predictions from the Monash and FRUITY models with the $s$-process observed abundance patterns of these objects, when extended to C, does not present major problems: the models generally predict [C/Fe] ratios higher than the observed value, allowing for the decrease due to the occurrence of the FDU. However, the five stars listed above show the common problem \citep[previously discussed by][]{cseh2022,denhartogh2023,vilagos2024} of the underproduction of the elements just beyond the first peak, especially Nb and Mo, in the models relative to the observations. In fact, three of them (HD~107541, HD~107270, and CPD$-64^{\circ}4333$) were previously selected as prominent examples of this problem, to be compared to models of the \textit{intermediate} neutron-capture ($i$-) process \citep[see Figure 14 of][]{denhartogh2023}. Furthermore, one of these three stars, HD~107541, also has the peculiarity of a relatively high $^{12}$C/$^{13}$C ratio. Therefore, it is plausible to propose that the process in the AGB star that resulted in the peculiar Nb and Mo abundances, may also have affected the carbon elemental and isotopic abundances. Although, the reverse is not true, as the other two stars with high C isotopic ratios do not share the high C, nor the Nb and Mo problem.

\section{Concluding remarks}\label{sec:conclusions}

Based on high-resolution spectroscopic data, we conducted a classical LTE analysis for a sample of 180 Ba giant stars, focusing on the light elements carbon, nitrogen, and oxygen. By fitting synthetic spectra to the observed features of the molecular C$_{2}$ band head at $\sim5\,635$~\AA\ and $^{12}$CN molecular band in the spectral range close to $7\,995 - 8\,005$~\AA, we derived abundances for carbon and nitrogen, respectively. The carbon isotopic ratios were derived from the $^{13}$CN features at $8\,004 - 8\,020$~\AA. Oxygen abundances, instead, were constrained from the parametrization reported by \citet[][]{melendez2002}, based on the [O\,{\sc i}] line. Our main observations are summarized below. 

The [C/Fe] ratios observed in the program stars range from $-0.30$ to $+0.60$~dex, increasing for lower metallicity regimes. As expected by AGB mass transfer, the C abundances in Ba stars are systematically larger than the values reported for normal giants. The trend with metallicity is similar to that of normal giant, as driven by the chemical evolution of the Galaxy. In the case of carbon, the fact that it is easier to enrich a star via AGB mass transfer at lower metallicity (both due to the more efficient TDU and the lower initial abundance) could also have contributed to the increasing trend. 

The nitrogen abundances display a flat behavior with metallicity, and similar abundances to the normal giant stars, indicative of the effect of the FDU. We found a moderate correlation with the [Na/Fe] ratios, also expected from the effect of the FDU. We further noticed a bifurcation below $\rm{[Fe/H]}\sim-0.30$~dex: while the [N/Fe] ratios tend do decrease for lower metallicities in normal giants, they are observed to increase in Ba stars. If this behavior is confirmed by larger statistics, it may be due to the activation of extra-mixing on the TP-AGB companion, converting some C into N as the metallicity decreases. The main outlier (BD$+09^{\circ}2384$) is a low metallicity star, with lower C and higher N than the general trend. The other outlier (HD~219116), also has nitrogen excess accompanied by a relatively low carbon abundance, as well as a low $^{12}\rm{C}/^{13}\rm{C}=10$. These two stars also point to the possible effect of extra-mixing on the AGB when $\rm{[Fe/H]}<-0.5$~dex. 

More than 80\% of the analyzed stars showed $^{12}\rm{C}/^{13}\rm{C}<20$, as expected by the FDU and extra-mixing on the RGB, although three stars (HD~66291, HD~39778, and HD~107541) shows higher ratios, $^{12}\rm{C}/^{13}\rm{C}>60$, which may reflect a stronger TP-AGB contribution.

As expected, we found a strong correlation between the carbon and the average $s$-process abundances in Ba stars, which is not observed in field giants, whose heavy element contents reflect the galactic chemical evolution. The entire sample present $\rm{C/O}<1$, as commonly found in Ba stars, except for one star (HD~107541), with $\rm{C/O}=1.32$, which may be placed to the family of the CH stars; this is the most C-rich and $s$-rich object of the sample. Despite the chemical puzzling features printed in HD~107541, the observed star of this binary system, with $M_{\rm{Ba}}\sim1.0$~M$_{\odot}$, is orbited by an evolved companion (white-dwarf) of $0.55$~M$_{\odot}$. It has a low-eccentricity ($\sim0.1$) and a period of $\sim3580$~days, thus lying close to other Ba stars in the eccentricity-period diagram \citep[e.g.,][]{escorza2023inp, escorza2023}. Another three stars (CPD$-64^{\circ}4333$, HD~24035, HD~204075, and HD~107270) have $\rm{C/O}$ close to unity. 

Finally, new models of red giant stars need to be calculated with an initial surface layer with the C and N composition corresponding to that of the material accreted from the AGB companion to quantitatively evaluate the effect of the FDU and potential extra-mixing processes in Ba stars. Models and scenarios invoked to explain the Nb and Mo overabundances in a fraction of Ba giants \citep{denhartogh2023} should incorporate the further constraints derived here from the C abundances. More broadly, a dedicated effort should be carried out to directly compare AGB models to Ba dwarfs, including improvement of the current sample, given that it derives from many different studies over several decades and it is not self-consistent as the Ba giant sample. \\

%% Also note that the akcnowlodgment environment does not support long amounts of text. If you have a lot of people and institutions to acknowledge, do not use this command. Instead, create a new \section{Acknowledgments}.

%\begin{acknowledgments}
This work has been developed under a fellowship of the PCI Program of the Ministry of Science, Technology and Innovation, financed by the Brazilian National Council of Research - CNPq. This study was financed in part by the Coordenação de Aperfeiçoamento de Pessoal de Nível Superior – Brasil (CAPES) – Finance Code 001. NAD acknowledges Funda\c{c}\~ao de Amparo \`a Pesquisa do Estado do Rio de Janeiro - FAPERJ, Rio de Janeiro, Brazil, for grant E-26/203.847/2022. NH acknowledges a fellowship (300466/2025-0) from the PCI Program - MCTI and CNPq, as well as financial support from FAPERJ through grant E-26/200.097/2025. ML and BC are supported by the Lend\"ulet Program LP2023-10 of the Hungarian Academy of Sciences. ML is also supported by the NKFIH Excellence Grant TKP2021-NKTA-64. We thank the referee for their thoughtful and constructive comments, which contributed to improving the manuscript. This work has made use of the {\sc vald} database, operated at Uppsala University, the Institute of Astronomy RAS in Moscow, and the University of Vienna. This research has made use of NASA’s Astrophysics Data System Bibliographic Services.
%\end{acknowledgments}


\vspace{5mm}
%\facilities{}
\software{{\sc iraf} \citep{tody1986}; {\sc moog} \citep{sneden1973, sneden2012}; R and RStudio \citep{rrstudio}}




\appendix
Atmospheric parameters and full abundance analysis performed for HD~26 are sumarized in Table~\ref{tab:app_hd26}. Table~\ref{tab:app_linelist} shows the line lists of the molecular/atomic transitions near the spectral region of the $\rm{C}_{2}$ and $\rm{CN}$ used in this work. Table~\ref{tab:app_linelist} is available in a machine-readable format. A portion is shown here for guidance regarding its form and content.

\section{HD~26}\label{sec:app_hd26}

\begin{table*}
    \centering
    \caption{Atmospheric parameters and full elemental abundances derived for HD\,26. For guidance, the solar abundances adopted in this work \citep[][]{grevesse1998}{}{}{} are listed in the second column. Stellar abundances in the scale $\log\epsilon$({\rm H}) = 12.0 and their standard deviations ($\sigma_{\rm obs}$), evaluated when three or more spectral lines were used, are provided in the third and fourth columns, respectively. Fifth column lists the number of lines used, otherwise the flag \textit{syn} indicates that abundances were derived from spectral synthesis; the number of lines used in the spectral synthesis are shown in parentheses. For oxygen, abundances were derived from parametrization of \citet[][MB02]{melendez2002}. Abundances in the [X/H] and [X/Fe] notations are shown in the sixth and seventh columns. At the end of the table, we give the carbon isotopic ratio.} \label{tab:app_hd26}
        \begin{tabular}{lccccccc}
        \toprule
        \multicolumn{8}{c}{HD\,26}\\
        \midrule
        \multicolumn{8}{c}{$T_{\rm{eff}}=5110$~K; $\log g=2.3$~(cm.s$^{-2}$); $\xi=1.8$~km.s$^{-1}$; $\rm{[Fe/H]}=-0.85$~dex}\\
        \midrule
        Species & $\log\epsilon_{\odot}$ & & $\log\epsilon$ &  $\sigma_{\rm obs}$ & $n$(\#) & [X/H] & [X/Fe] \\
        \midrule 
        C\,(C$_{2}$ 5\,635)  & 8.52 & & 8.24    &  --- & syn(1) & $-$0.28 & $+$0.57 \\
        N                    & 7.92 & & 7.70    &  --- & syn(1) & $-$0.22 & $+$0.63 \\
        O\,{\sc i}           & 8.83 & & 8.21    &  --- & MB02   & $-$0.62 & $+$0.23 \\
        Na\,{\sc i}          & 6.33 & & 5.63    & 0.07 & 03     & $-$0.70 & $+$0.15\\
        Mg\,{\sc i}          & 7.58 & & 7.16    & 0.15 & 03     & $-$0.42 & $+$0.43 \\
        Al\,{\sc i}          & 6.47 & & 5.77    & 0.11 & 03     & $-$0.70 & $+$0.15 \\
        Si\,{\sc i}          & 7.55 & & 6.94    & 0.07 & 04     & $-$0.61 & $+$0.24 \\
        Ca\,{\sc i}          & 6.36 & & 5.61    & 0.07 & 08     & $-$0.75 & $+$0.10 \\
        Ti\,{\sc i}          & 5.02 & & 4.27    & 0.05 & 06     & $-$0.75 & $+$0.10 \\
        Cr\,{\sc i}          & 5.67 & & 4.66    & 0.02 & 03     & $-$1.01 & $-$0.16 \\
        Fe\,{\sc i}          & 7.50 & & 6.65    &  --- & 53     & $-$0.85 &    ---  \\
        Fe\,{\sc ii}         & 7.50 & & 6.65    &  --- & 10     & $-$0.85 &    ---  \\
        Ni\,{\sc i}          & 6.25 & & 5.45    & 0.11 & 17     & $-$0.80 & $+$0.05 \\
        Rb\,{\sc i}          & 2.60 & & 2.10    &  --- & syn(1) & $-$0.50 & $+$0.35 \\
        Sr\,{\sc i}          & 2.97 & & 3.98    &  --- & syn(1) & $+$1.01 & $+$1.86 \\
        Y\,{\sc ii}          & 2.24 & & 2.60    & 0.15 & 04     & $+$0.36 & $+$1.21 \\
        Zr\,{\sc ii}         & 2.60 & & 3.10    & 0.29 & 03     & $+$0.50 & $+$1.35 \\
        Nb\,{\sc i}          & 1.42 & & 2.42:   &  --- & syn(1) & $+$1.00 & $+$1.85 \\
        Mo\,{\sc i}          & 1.92 & & 2.59    &  --- & 02     & $+$0.67 & $+$1.52 \\
        Ru\,{\sc i}          & 1.84 & & 3.25    &  --- & 01     & $+$1.41 & $+$2.26 \\
        La\,{\sc ii}         & 1.17 & & 1.77    & 0.12 & 04     & $+$0.60 & $+$1.45 \\
        Ce\,{\sc ii}         & 1.58 & & 2.42    & 0.11 & 07     & $+$0.84 & $+$1.69 \\
        Nd\,{\sc ii}         & 1.50 & & 2.27    & 0.12 & 13     & $+$0.77 & $+$1.62 \\ 
        Sm\,{\sc ii}         & 1.01 & & 1.56    & 0.08 & 08     & $+$0.55 & $+$1.40 \\ 
        Eu\,{\sc ii}         & 0.51 & & 0.41    &  --- & syn(1) & $-$0.10 & $+$0.75 \\        
         W\,{\sc i}          & 0.65 & & 2.11:   &  --- & syn(1) & $+$1.46 & $+$2.31 \\  
         \midrule
         \multicolumn{8}{c}{$\rm{C}^{12}/\rm{C}^{13}=16$}\\
        \bottomrule
    \end{tabular}
\end{table*}

\section{Line list}\label{sec:app_linelist}

\begin{table}
    \centering
    \caption{Line list near the spectral region of the $\rm{C}_{2}$ and $\rm{CN}$ molecular absorption features used in this work. The molecules/atoms are identified in the first column. The wavelength, excitation potential (E.P.), $\log$~\textit{gf} values, and dissociation energy (D.E.) are listed in the columns 2, 3, 4, and 5, respectively. This table is fully available in machine-readable format. A small portion is shown here for guidance regarding its form and content. }\label{tab:app_linelist}
    \begin{tabular}{lcccc}
    \toprule
           Molecule/                 &  Wavelength &  E.P.   & $\log$\textit{gf} &  D.E.  \\
            Atom                     &  (\AA)      &  (eV)   &                   &  (eV)  \\
    \midrule
    \multicolumn{5}{c}{$\rm{C}_{2}$ region at 5628 - 5638~\AA}                            \\
    \midrule
           $^{12}\rm{C}^{13}\rm{C}$  &  5628.001   &  0.232  & -6.365            &  6.244 \\
           $^{12}\rm{C}^{13}\rm{C}$  &  5628.007   &  0.227  & -5.988            &  6.244 \\
           $^{12}\rm{C}^{12}\rm{C}$  &  5628.025   &  0.330  & -0.597            &  6.270 \\
           $^{12}\rm{C}^{13}\rm{C}$  &  5628.027   &  0.573  & -4.486            &  6.244 \\
           $^{12}\rm{C}^{13}\rm{C}$  &  5628.1031  &  0.0426 & -3.502            &  7.724 \\
           ...                       &   ...       &  ...    & ...               &  ...   \\
    \midrule
           \multicolumn{5}{c}{$\rm{CN}$ region at 7997 - 8008~\AA}                       \\
    \midrule
           $^{12}\rm{C}^{14}\rm{N}$  &  7997.132   &  2.395  & -5.046            & 7.724 \\
           $^{12}\rm{C}^{14}\rm{N}$  &  7997.291   &  1.334  & -3.268            & 7.724 \\
           Fe\,{\sc i}               &  7997.3012  &  4.1426 & -3.718            &       \\
           $^{12}\rm{C}^{14}\rm{N}$  &  7997.396   &  1.356  & -3.460            & 7.724 \\ 
           $^{12}\rm{C}^{14}\rm{N}$  &  7997.746   &  1.417  & -2.860            & 7.724 \\
           ...                       &   ...       &  ...    & ...               &  ...   \\       
        \bottomrule
    \end{tabular}
\end{table}


\bibliography{paper_cno_barium_stars_revised_R2}{}
\bibliographystyle{aasjournal}



\end{document}